\documentclass[titlepage, 12pt]{article}

\usepackage{natbib}
\usepackage{amsmath, amsbsy, amsthm, epsfig, epsf, psfrag, booktabs, graphicx, amssymb, enumerate}
\usepackage{hyperref}
\usepackage{hypernat}

\usepackage[margin=1in]{geometry}
\usepackage{rotating} %, endrotfloat}

\usepackage{authblk}
\usepackage{rotating}
\usepackage{multicol}

\newcommand{\dif}{\mathrm{d}}%
\newcommand{\Var}{\mathrm{Var}}%
\newcommand{\E}{\mathrm{E}}%
\begin{document}

\title{Fast Accelerated Failure Time Modeling for Case-Cohort Data}

\renewcommand\Affilfont{\normalfont\small}
\author[1]{Sy Han Chiou}
\author[1]{Sangwook Kang}
\author[1,2,3]{Jun Yan}
\affil[1]{Department of Statistics, University of Connecticut}
\affil[2]{Institute for Public Health Research, University of Connecticut Health Center}
\affil[3]{Center for Environmental Sciences \& Engineering, University of Connecticut}
% \date{}

\maketitle

\begin{abstract}
Semiparametric accelerated failure time (AFT) models directly relate
the predicted failure times to covariates and are a useful alternative
to models that work on the hazard function or the survival function.
For case-cohort data, much less development has been done with AFT models.
In addition to the missing covariates outside of the sub-cohort in controls,
challenges from AFT model inferences with full cohort are retained.
The regression parameter estimator is hard to compute because the
most widely used rank-based estimating equations are not smooth.
Further, its variance depends on the unspecified error distribution,
and most methods rely on computationally intensive bootstrap to estimate it.
We propose fast rank-based inference procedures for AFT models, applying
recent methodological advances to the context of case-cohort data.
Parameters are estimated with an induced smoothing approach that smooths
the estimating functions and facilitates the numerical solution.
Variance estimators are obtained through efficient resampling methods
for nonsmooth estimating functions that avoids full blown bootstrap.
Simulation studies suggest that the recommended procedure provides
fast and valid inferences among several competing procedures.
Application to a tumor study demonstrates the utility of the proposed
method in routine data analysis.

\end{abstract}

\section{Introduction}
\label{sect:intr}

A case-cohort design \citep{Pren:case:1986} is an effective and economical
design which reduces the effort and cost of a full-scale cohort study.
Such design originated to allow efficient analysis of studies where it is
too expensive and time consuming to collect and analyze data on all subjects.
Cases and controls refer to subjects who have and have not, respectively,
developed the disease of interest by the end of the study period.
A case-cohort design is typically composed of two steps.
First, a subset called sub-cohort is randomly selected
from the whole cohort regardless of their disease status.
Second, the remaining cases in the cohort are added to the sub-cohort.
Cases and controls refer to subjects who have and have not, respectively,
developed the disease of interest by the end of the study period.
Measurement on the main risk factors are taken only on subjects in the
sub-cohort and the remaining cases outside of the sub-cohort.
This leads to substantial reduction in the effort and cost of conducting
large scale cohort studies, especially when the disease of interest is
rare or the main risk factors are expensive to measure.

A semiparametric accelerated failure time (AFT) model is a log-linear
model for the failure times with unspecified error distribution.
It directly relates the failure time to covariates such that the effect
of a covariate is to multiply the predicted failure time by a constant.
For failure time data from case-cohort studies, most statistical
methods have focused on semiparametric models that work on either
the hazard function \citep{Pren:case:1986, Self:Pren:asym:1988,
Lin:Ying:cox:1993, Barl:robu:1994, Ther:Li:comp:1999,
Kuli:Lin:addi:2000, Sun:Sun:Flou:addi:2004, Kang:Cai:marg:2009},
or the survival function \citep{Chen:weig:2001, Chen:fitt:2001,
Kong:Cai:Sen:weig:2004, Lu:Tsia:semi:2006}.
Parametric AFT models were considered by \citet{Kalb:Lawl:like:1988}.
Inferences about semiparametric AFT models for case-cohort data
are much less developed, with only a few recent works
\citep{Nan:Yu:Kalb:cens:2006, Yu:Wong:Yu:buck:2007,
Kong:Cai:case:2009, Yu:buck:2011}.

Inferences for semiparametric AFT models have been difficult
for not only case-cohort data but also for complete data.
The most important estimator is the rank-based estimator motivated from
inverting the weighted log-rank test \citep{Pren:line:1978}, with asymptotic
properties rigorously studied \citep{Tsia:esti:1990, Ying:larg:1993}.
Nevertheless, the estimator has not been as widely used as it should
be due to lack of efficient and reliable computing algorithm to
obtain both parameter estimates and their standard errors.

The parameter estimates are hard to compute because the
most widely used rank-based estimating equations are not smooth.
Recent works shed light on bringing AFT models into routine
data analysis practice, including case-cohort studies.
\citet{Jin:Lin:Wei:Ying:rank:2003} exploited that the rank-based
estimating equation with Gehan's weight is the gradient of an objective
function and obtained estimates by solving it with linear programming.
This approach was adapted to case-cohort data by \citet{Kong:Cai:case:2009}.
Nevertheless, the optimization with linear programming is still
computationally very demanding, especially for larger sample sizes.
A more computing efficient approach for rank-based inference is
the induced smoothing procedure of \citet{Brow:Wang:indu:2007}.
This approach is an application of the general induced smoothing
method of \citet{Brow:Wang:stan:2005}, where the discontinuous
estimating equations are replaced with a smoothed version, whose
solutions are asymptotically equivalent to those of the former.
The smoothed estimating equations are differentiable, thus
facilitates rapid numerical solution.

Direct estimation of the variance is difficult because it involves
nonparametric estimation of the unspecified error distribution.
Most existing methods rely on bootstrap which is very computing intensive.
\citet{Jin:Lin:Wei:Ying:rank:2003} estimated the variance through
a multiplier resampling method, which requires a large bootstrapping
sample in order to obtain a reliable variance estimate.
For case-cohort data, \citet{Kong:Cai:case:2009} adopted a specially
designed bootstrap procedure \citep{Wach:Gail:Pee:Broo:altn:1989}.
The demanding computing task in linear programming is amplified
because it requires solving estimating equations for each bootstrap sample.
\citet{Huan:cali:2002} proposed an easy-to-compute variance estimator
based on the asymptotic linearity property of the estimating equations.
A decomposition matrix of the variance matrix is estimated by solving
estimating equations, but the number of the estimating equations to
solve is much smaller; it is just the dimension of the parameters.
For general nonsmooth estimating functions, \citet{Zeng:Lin:effi:2008}
proposed a resampling strategy that does not require solving
estimating equations or minimizing objective functions.
Instead, it only involves evaluations of estimating functions
and simple linear regression in estimating the slope matrix.
The resulting variance estimators are computationally more
efficient and stable than those from existing resampling methods.

In this article, we propose a fast rank-based inference procedure for
semiparametric AFT models in the context of case-cohort studies.
The parameters are estimated with an induced smoothing approach.
Variance estimators are obtained through an efficient resampling methods
for nonsmooth estimating functions that avoids full blown bootstrap.
Of course, the methods also apply to full cohort data.

The rest of this article is organized as follows.
Point estimation procedures based on smoothed estimating equations
for case-cohort data when the sub-cohort is a simple random sample
from the full cohort are proposed in Section~\ref{sect:poin}.
Four variance estimation procedures, one based on full multiplier
bootstrap and three based on possibly multiplier bootstrap-aided
sandwich variance estimator, are proposed in Section~\ref{sect:vari}.
A large scale simulation study is reported in Section~\ref{sect:simu},
comparing the performances of the variance estimator and their timings.
The methods are applied to a tumor study with both case-cohort
data and full cohort data in Section~\ref{sect:appl}.
A discussion concludes in Section~\ref{sect:disc}.

\section{Point Estimation}
\label{sect:poin}

Let $\{T_i, C_i, X_i\}$, $i = 1, \ldots, n$, be $n$
independent copies of $\{T, C, X\}$, where
$T_i$ and $C_i$ are log-transformed failure time and log-transformed
censoring time, $X_i$ is a $p\times 1$ covariate vector,
and given $X$, $C$ and $T$ are assumed to be independent.
A semiparametric AFT model has the form
\begin{equation*}
  T_i = X_i^{\top} \beta + \epsilon _i, \qquad i = 1, \ldots, n,
\end{equation*}
where $\beta$ is an unknown $p \times 1$ vector of regression
parameters, $\epsilon_i$'s are independent and identically
distributed random variables with an unspecified distribution.
It is also assumed that $\epsilon_i$'s are independent of $X_i$.

In a full cohort study, due to censoring, the observed data are
$(Y_i, \Delta_i, X_i)$, $i = 1, \ldots, n$,
where $Y_i = \min(T_i, C_i)$, $\Delta_i = I[T_i < C_i]$, and 
$I[\cdot]$ is the indicator function.
A rank based estimating equation with Gehan's weight is
\begin{equation}
  U_n(\beta)=\sum_{i=1}^n \sum_{j=1}^n \Delta_i (X_i- X_j) I[e_j(\beta) \geq e_i(\beta)] = 0,
  \label{eq:Un}
\end{equation}
where $e_i(\beta) = Y_i - X_i^\top\beta$.
The root of~\eqref{eq:Un} is consistent to the true parameter
$\beta_0$, and is asymptotically normal \citep{Tsia:esti:1990}.
Despite these nice properties, even for the most promising
method to date that solves it via linear programming
\citep{Jin:Lin:Wei:Ying:rank:2003}, the computing burden
increases drastically when bootstrapping is used to estimate
the variance of the estimator.

For a case-cohort study, the covariate vector $X_i$'s
are not completely available for each individual.
Measurement of some covariates is taken only on the subjects
in the sub-cohort and cases outside the sub-cohort, and, thus,
estimating function~\eqref{eq:Un} cannot be evaluated.
Using the observed data naively in~\eqref{eq:Un} would
lead to misleading results because the case-cohort sample
is biased --- it includes all cases but only a fraction of controls.
It is possible, however, to adjust the biases by incorporating a
weight that depends on the selection scheme of case-cohort samples.
Suppose we select a sub-cohort of size $\tilde n$ by simple
random sampling without replacement from the whole cohort.
Let $\xi_{i}$ be the sub-cohort indicator; $\xi_i=1$ if the $i$th
observation is in the sub-cohort and $\xi_i = 0$ otherwise.
Let $p = \lim_{n\to\infty} p_n$, where $p_n = \tilde n / n$
is the sub-cohort inclusion probability.
Under these assumptions, the desired case-cohort weight is
$h_{i} = \Delta_{i} + (1-\Delta_{i})\xi_{i} / p_n$.
The weight-adjusted estimating equation~\eqref{eq:Un} becomes
\begin{equation}
  U_n^c(\beta) = \sum_{i=1}^n \sum_{j=1}^n h_j \Delta_i (X_i- X_j) I[e_j(\beta)\geq e_i(\beta)] = 0.
  \label{eq:UnC}
\end{equation}
The solution to~\eqref{eq:UnC}, $\hat{\beta}_n$, remains to be
consistent and asymptotically normal \citep{Kong:Cai:case:2009}.

For full cohort data, a computationally more efficient
approach for rank-based inference with Gehan's weight is
the induced smoothing procedure of \citet{Brow:Wang:indu:2007}.
Such smoothing method leads to continuously differentiable
estimating equations that can be solved with standard numerical methods.
Let $Z$ be a $p$-dimensional standard normal random vector.
The estimating function $U_n(\beta)$ in~\eqref{eq:Un}
is replaced with $E[U_n(\beta + n^{-1/2}Z)]$, where
the expectation is taken with respect to $Z$.
This lead to
\begin{equation}
  \label{eq:UnSmooth}
  \tilde{U}_n(\beta) = \sum_{i=1}^n \sum_{j=1}^n \Delta_i (X_i- X_j) \Phi \left[ \frac{e_j(\beta)-e_i(\beta)}{r_{ij}^2} \right] = 0,
\end{equation}
where $r_{ij}^2 = n^{-1} (X_i - X_j)^\top(X_i - X_j)$ and $\Phi(\cdot)$
denotes the standard normal cumulative distribution function.
The solution to~\eqref{eq:UnSmooth} is consistent to
$\beta_0$ and has the same asymptotic distribution as the
solution to~\eqref{eq:Un} \citep{John:Stra:indu:2009}.

For case-cohort data, we propose a smoothed version of~\eqref{eq:UnC}
by adapting the idea of \citet{Brow:Wang:indu:2007}.
Specifically, we replace $U^c_n(\beta)$ with $E[U^c_n(\beta + n^{-1/2}Z)]$ to
obtain the induced smooth version of~\eqref{eq:UnC},
\begin{equation}
  \label{eq:UnSmoothC}
  \tilde{U}_n^c(\beta) = E[U^c_n(\beta+n^{-1/2}Z)]= \sum_{i=1}^n \sum_{j=1}^n h_j \Delta_i (X_i- X_j) \Phi \left[ \frac{e_j(\beta)-e_i(\beta)}{r_{ij}^2} \right].
\end{equation}
The solution $\tilde{\beta}_n$ to~\eqref{eq:UnSmoothC} is a
consistent estimator to $\beta_0$ and is asymptotically normal.
Furthermore, the asymptotic distribution of $\tilde{\beta}_n$
is also the same as that of $\hat{\beta}_n$.
These arguments can be justified similarly as those in
\citet{John:Stra:indu:2009}.

\section{Variance Estimation}
\label{sect:vari}

The asymptotic variance of $\tilde\beta_n$ is even harder to
estimate for case-cohort data than for full cohort data
because of the extra complexity caused by the data structure.
The terms in the summation in $\tilde{U}_n(\beta)$ are not independent
since the sub-cohort is drawn from the full cohort without replacement.
We propose four variance estimators; one is fully resampling based
while the other three use resampling to a component of the sandwich
variance estimator.

\subsection{Multiplier Bootstrap}

The multiplier bootstrap estimator of \citet{Jin:Lin:Wei:Ying:rank:2003}
is adapted to case-cohort data by inserting proper case-cohort weights,
$h_i$'s, in the multiplier bootstrap estimating equations.
Let $\eta_i$, $i = 1, \dots, n$, be independent and
identically distributed positive random variables with
$\E(\eta_i) = \Var(\eta_i) = 1$.
Define
\begin{equation}
  \tilde{U}_n^{c*}(\beta) = \sum_{i=1}^n \sum_{j=1}^n \eta_i\eta_jh_j\Delta_i (X_i- X_j) \Phi \left[ \frac{e_j(\beta)-e_i(\beta)}{r_{ij}^2} \right].
  \label{eq:UnSmoothC:Res}
\end{equation}
For a realization of $(\eta_1, \ldots, \eta_n)$, the solution
to~\eqref{eq:UnSmoothC:Res} provides one draw of
$\tilde\beta_n$ from its asymptotic distribution.
By repeating this process a large number $B$ times, the variance
matrix of $\tilde\beta_n$ can be estimated directly by the sampling
variance matrix of the bootstrap sample of $\tilde\beta_n$.

Since the asymptotic variance of $\hat{\beta}_n$ is the same as that
of $\tilde{\beta}_n$, the covariance matrix of $\tilde{\beta}_n$ can
also be estimated by~\eqref{eq:UnC} through multiplier bootstrap.
This is, however, not recommended because it would need to
solve a large number $B$ nonsmooth estimating equations.
As will be seen in our simulation study, even with the
computationally more efficient smoothing estimating equations,
the multiplier bootstrap approach can still be very time consuming,
especially for larger sample sizes or more covariates.

\subsection{Sandwich Estimator}

To improve the computational efficiency, we consider alternative
variance estimation procedures based on the sandwich form
that avoid solving estimating equations repetitively.
The asymptotic variances of $\hat\beta_n$ and
$\tilde\beta_n$ are the same, both having a sandwich form.
Under some regularity conditions \citep{Zeng:Lin:effi:2008},
uniformly in a neighborhood of $\beta_0$,
equation~\eqref{eq:UnC} can be expressed as
\begin{equation*}
  n^{-1/2}U^c_n(\beta)= n^{-1/2}\sum_{i=1}^nh_iS_i(\beta_0) + An^{1/2}(\beta-\beta_0) + o_p(1 + n^{1/2}\| \beta - \beta_0 \|),
\end{equation*}
where $S_i(\beta_0)$ is a zero-mean random vector,
and $A$ is asymptotic slope matrix of $n^{-1/2}\tilde{U}^c_n(\beta_0)$.
The analytical details of $S_i(\beta_0)$ for case-cohort
data is presented in the Appendix.
The asymptotic variance matrix of
$\sqrt{n}(\tilde\beta_n - \beta_0)$
is $n\Sigma = nA^{-1}V(A^{-1})^\top$,
where $V$ is the variance  of $n^{-1/2}\sum_{i=1}^nh_{i}S_i(\beta_0)$.
Estimation of $\Sigma$ involves estimating $V$ and $A$
by estimator $V_n$ and $A_n$, respectively.
The variance estimator then has the sandwich form
$\hat{\Sigma}_n = A_n^{-1} V_n (A_n^{-1})^\top$.

\subsubsection{Estimation of $V$}
Matrix $V$ can be estimated either through a closed-form
estimator or through bootstrapping the estimating equations.
For case-cohort data, due to the correlated feature of $\xi_{i}$'s
in $h_{i}$'s, $V$ is different from its full cohort counterpart.
There are two sources of variations contributing to $V$:
variation due to the sampling of a full cohort ($V_{1}$)
and variation due to the sampling of a sub-cohort
within the full cohort ($V_{2}$).
In particular, we have
\begin{equation*}
  V = V_{1} + \frac{1-p}{p}V_{2} = E \left[S_{i}(\beta_0)S_{i}(\beta_0)^{\top}\right] + \frac{1-p}{p} \Var\left[(1 - \Delta_i)S_{i}(\beta_0)\right],
\end{equation*}
where $V_2$ vanishes if full cohort data are available.

\paragraph{Closed-form}
With explicit expressions for $S_i(\beta)$'s in the Appendix,
a closed-form estimator of $V$ is
\begin{equation*}
V_n = V_{1n} +  \frac{1-p_n}{p_n}V_{2n}
\end{equation*}
where
\begin{equation*}
  V_{1n} = n^{-1}\sum_{i=1}^{n}h_i \hat{S}_{i}(\hat\beta_n) \hat{S}_{i}^\top(\hat\beta_n),
\end{equation*}
and
\begin{equation*}
  V_{2n} = n^{-1}\sum_{i=1}^{n}h_i(1-\Delta_{i}) \hat{S}_{i}(\hat\beta_n) \hat{S}_{i}^\top(\hat\beta_n) - \left\{n^{-1}\sum_{i=1}^{n}h_i(1-\Delta_{i})\hat{S}_{i}(\hat\beta_n)\right\}\left\{n^{-1}\sum_{i=1}^{n}h_i(1-\Delta_{i})\hat{S}_{i}(\hat\beta_n)\right\}^{\top},
\end{equation*}
and $\hat{S}_{i}(\hat\beta_n)$ is obtained by replacing unknown
quantities in $S_{i}(\beta)$ with their sample counterparts.

\paragraph{Multiplier Bootstrap}
When $\hat{S}_i(\hat\beta_n)$ have complicated expressions,
it is more convenient and perhaps more accurate to estimate
$V$ via bootstrap \citep{Zeng:Lin:effi:2008}.
Because $U_n^c$ and $\tilde U_n^c$ have the same asymptotic
distribution, we apply the multiplier bootstrap approach to $\tilde U_n^c$.
Evaluation of~\eqref{eq:UnSmoothC:Res} at $\hat\beta_n$ with
each realization of $(\eta_1, \ldots, \eta_n)$ provides one
bootstrap replicate of $\tilde{U}_n^{c*}(\hat\beta_n)$.
With $B$ replicates, we estimate $V$ by the sample variance
of the bootstrap sample of $\tilde{U}_n^{c*}(\hat\beta_n)$.
The bootstrap here is much less demanding than the full
multiplier bootstrap above, because it only involves evaluations
of estimating equations instead of solving them to obtain each
bootstrap replicate.

\subsubsection{Estimation of $A$}

With $V$ estimated by $V_n$, we next propose three
approaches to estimate the slope matrix $A$.
Depending whether $V_n$ is based on closed-form or multiplier
bootstrap, we will have two versions of estimator of $\Sigma$
for each approach of slope matrix estimation.

\paragraph{Induced Smoothing}
With $\tilde U_n^c$, the smoothed version of $U_n^c$,
the slope matrix $A$ can be estimated directly by
\begin{equation*}
  A_n = \frac{1}{n} \frac{\partial}{\partial \beta^{\top}}
  \tilde U_n^c(\hat\beta_n).
\end{equation*}
The close-form expression of $A_n$ can be evaluated easily.
The variance estimator then has the sandwich form
$\hat{\Sigma}_n = A_n^{-1} V_n (A_n^{-1})^\top$.

\paragraph{Smoothed Huang's (2002) Approach}
\citet{Huan:cali:2002} avoided the difficulty in estimating
the slope matrix of nonsmooth estimating equations by
exploiting the asymptotic linearity of the estimating equations.
Nevertheless, this approach still requires solving $p$ nonsmooth
estimating equations, whose convergence may be a problem.
We adapt Huang's approach by replacing the $p$ nonsmooth estimating
equations with their smoothed versions.
Let $V_n = L_n^{\top} L_n$ be the Cholesky decomposition of $V_n$.
Let $q_{nj}$ be the solution to the following estimating equations
for $\gamma$, $j = 1, \ldots, p$,
\begin{equation*}
  n^{-1} \tilde{U}^c_n( \gamma ) = n^{-1/2} l_j,
\end{equation*}
where $l_j$ is the $j$th column of $L_n$.
The solutions can be obtained with from general purpose nonlinear
equation solvers; in our implementation we used R packages
nleqslv \citep{Rpkg:nleqslv} and BB \citep{Vara:Gilb:BB:2009}.
Let $Q_n$ be the matrix whose $j$th column is $q_{nj} - \hat\beta_n$.
Then $Q_n^{\top}Q_n$ is an estimate of $\Sigma$.

With the adaptation to smooth estimating equations, this approach
has an advantage compared to the induced smoothing approach in that
the closed-form derivative matrix is not required, and, hence, can
be applied to more general nonsmooth estimating equations.

\paragraph{Zeng and Lin's (2008) Approach}
\citet{Zeng:Lin:effi:2008} proposed to estimate the slope matrix by
regressing the perturbed estimating functions on the perturbations.
Let $Z_b$, $b = 1, \ldots, B$, be $B$ realizations
of a $p$-dimensional standard normal random vector.
For case-cohort data, let $U_{nj}^c$ be the $j$th component of $U_{n}^c$.
We estimate the $j$th row of $A$, $j = 1, \ldots, p$,
by $A_{nj}$, the least squares estimate of the regression
coefficients when regressing $n^{-1/2} U_{nj}(\hat\beta_n + n^{-1/2} Z_b)$
on $Z_b$, $i = 1, \ldots, n$.
The variance estimator also has the sandwich form
$\hat{\Sigma}_n = A_n^{-1} V_n (A_n^{-1})^\top$.

This approach differs from the induced smoothing approach
in that the slope matrix $A$ is estimated via a resampling
procedure that involves $p$ least squares regressions,
instead of taking the derivatives of a smooth function.
It can be viewed as an empirical version of the induced
smoothing approach.

\section{Simulation}
\label{sect:simu}
We conducted an extensive simulation study to assess the
performance of the our point and variance estimators.
Failure time $T$ was generated from AFT model
\begin{equation*}
  \log(T) = 2 + X_{1} + X_{2} + X_{3} + \epsilon,
\end{equation*}
where $X_{1}$ was Bernoulli with rate 0.5, $X_{2}$ and
$X_{3}$ were uncorrelated standard normal variables.
Censoring time $C$ was generated from $\mathrm{unif}(0, \tau)$
where $\tau$ was tuned to achieve desired censoring rate $C_p$.
The distribution of $\epsilon$ had three types:
standard normal, standard logistic, or standard Gumbel,
abbreviated by N, L, and G, respectively.
The censoring rate $C_p$ had two levels, 90\% and 97\%, representing
a mildly rare disease and a very rare disease, respectively.
For the mildly rare disease, the full cohort size was set to be
1500 and the case-cohort size was set to $\bar m = 300$ on average.
For the very rare disease, the full cohort sizes were set to be 1500 and
3000, each with case-cohort sizes averaged at $\bar m \in \{150, 300\}$.
The sub-cohort sampling proportion $p_n$ was set to yield the desired
average case-cohort size given censoring rate and full cohort size.
For each viable combination, we generated 1000 datasets.

Given a dataset, point estimates of regression coefficients were
obtained from both nonsmooth and smoothed estimating equations.
The estimator from the nonsmooth version was obtained using
linear programming \citep{Jin:Lin:Wei:Ying:rank:2003}, denoted by LP.
The estimator from the induced smoothing approach with
estimating equations~\eqref{eq:UnSmoothC} was obtained using
R package nleqslv \citep{Rpkg:nleqslv}, denoted by IS.
The two estimators are expected to be asymptotically the same,
but with the IS estimator obtained much faster.
Eight variance estimates were computed for the point estimate.
The first two were full multiplier bootstrap estimates, denoted by
MB, one based on the LP approach and the other based on the IS approach.
The rest six were sandwich estimates constructed by combinations
of three approaches to estimate $A$ and two approaches to estimate $V$.
We use abbreviations IS, SH, and ZL to denote the induced smoothing,
smoothed Huang's, and Zeng and Lin's approach for $A$, respectively.
We use abbreviations CF and MB to denote the closed-form estimate
approach and the multiplier bootstrap approach for $V$, respectively.

\begin{sidewaystable}[tbp]
  \caption{Summary of simulation results based on 1000 replications
    for full cohort size 1500 and censoring rate 90\%.
    The bootstrapping size is 500 for each replication.
    PE is average of point estimates;
    ESE is the empirical standard deviation of the parameter estimates;
    ASE is the average of the standard error of the estimator;
    CP is the coverage percentage of 95\% confidence interval.}
  \label{tab:1500:90}
  \begin{center}
    \renewcommand\tabcolsep{2pt}
    \begin{tabular}{ll rr l rr l rrrrrrrr l rrrrrrrr}
      \midrule
      Error & $\beta$ & \multicolumn{2}{c}{PE}&& \multicolumn{2}{c}{ESE}&& \multicolumn{8}{c}{ASE}&& \multicolumn{8}{c}{CP(\%)}\\
      \cmidrule(lr){3-4}\cmidrule(lr){6-7}\cmidrule(lr){9-16}\cmidrule(lr){18-25}
&& LP & IS && LP & IS && \multicolumn{2}{c}{MB} & \multicolumn{2}{c}{IS} & \multicolumn{2}{c}{SH} & \multicolumn{2}{c}{ZL} &&  \multicolumn{2}{c}{MB} & \multicolumn{2}{c}{IS} & \multicolumn{2}{c}{SH} & \multicolumn{2}{c}{ZL}\\
     \cmidrule(lr){8-10}\cmidrule(lr){11-12}\cmidrule(lr){13-14}\cmidrule(lr){15-17}\cmidrule(lr){18-19}\cmidrule(lr){20-21}\cmidrule(lr){22-23}\cmidrule(lr){24-25}
     &&& && &&& LP & IS& CF& MB & CF & MB & CF & MB && LP & IS& CF & MB & CF & MB & CF & MB\\
      \midrule
      N & $\beta_1$ & 0.997 & 1.000 &  & 0.170 & 0.170 &  & 0.161 & 0.161 & 0.161 & 0.163 & 0.155 & 0.157 & 0.161 & 0.163 &  & 93.4 & 93.8 & 94.0 & 94.2 & 93.0 & 93.5 & 94.0 & 94.5 \\
      & $\beta_2$ & 1.004 & 1.009 &  & 0.090 & 0.090 &  & 0.089 & 0.089 & 0.089 & 0.090 & 0.086 & 0.087 & 0.089 & 0.090 &  & 93.9 & 93.8 & 94.0 & 94.3 & 93.0 & 93.5 & 94.0 & 94.1 \\
      & $\beta_3$ & 1.000 & 1.004 &  & 0.093 & 0.093 &  & 0.089 & 0.088 & 0.089 & 0.090 & 0.102 & 0.103 & 0.089 & 0.090 &  & 93.8 & 94.0 & 94.1 & 94.3 & 96.6 & 96.7 & 94.0 & 94.6 \\
      L & $\beta_1$ & 0.998 & 1.000 &  & 0.284 & 0.284 &  & 0.274 & 0.274 & 0.273 & 0.275 & 0.269 & 0.271 & 0.273 & 0.275 &  & 94.4 & 94.5 & 94.6 & 94.9 & 93.9 & 94.2 & 94.4 & 94.7 \\
      & $\beta_2$ & 1.008 & 1.011 &  & 0.150 & 0.150 &  & 0.149 & 0.149 & 0.148 & 0.149 & 0.148 & 0.149 & 0.148 & 0.149 &  & 94.9 & 94.9 & 94.6 & 94.6 & 94.1 & 94.8 & 94.7 & 94.9 \\
      & $\beta_3$ & 1.012 & 1.015 &  & 0.151 & 0.151 &  & 0.149 & 0.149 & 0.149 & 0.150 & 0.159 & 0.160 & 0.149 & 0.150 &  & 94.2 & 94.4 & 94.0 & 94.0 & 95.9 & 96.1 & 93.8 & 94.6 \\
      G & $\beta_1$ & 0.999 & 1.003 &  & 0.148 & 0.148 &  & 0.142 & 0.143 & 0.143 & 0.145 & 0.137 & 0.139 & 0.143 & 0.145 &  & 94.7 & 94.9 & 94.4 & 94.5 & 93.1 & 93.2 & 94.4 & 94.7 \\
      & $\beta_2$ & 0.999 & 1.004 &  & 0.082 & 0.082 &  & 0.079 & 0.079 & 0.079 & 0.080 & 0.075 & 0.077 & 0.079 & 0.080 &  & 93.2 & 93.6 & 93.6 & 94.1 & 91.9 & 92.3 & 93.6 & 94.2 \\
      & $\beta_3$ & 1.001 & 1.006 &  & 0.082 & 0.082 &  & 0.079 & 0.079 & 0.079 & 0.081 & 0.093 & 0.094 & 0.079 & 0.081 &  & 93.3 & 92.7 & 93.3 & 93.6 & 97.3 & 97.2 & 93.2 & 94.2 \\
      \bottomrule
    \end{tabular}
  \end{center}
\end{sidewaystable}

\begin{sidewaystable}[tbp]
  \caption{Summary of simulation results based on 1000 replications for
    full cohort size 3000 and and censoring rate 97\%.
    The bootstrapping size is 500 for each replication.
    PE is average of point estimates;
    ESE is the empiricalstandard deviation of the parameter estimates;
    ASE is the average of the standard error of the estimator;
    CP is the coverage percentage of 95\% confidence interval.}
  \label{tab:3000:97}
  \begin{center}
    \renewcommand\tabcolsep{2pt}
    \begin{tabular}{ll rr l rr l rrrrrrrr l rrrrrrrr}
      \midrule
      Error & $\beta$ & \multicolumn{2}{c}{PE}&& \multicolumn{2}{c}{ESE}&& \multicolumn{8}{c}{ASE}&& \multicolumn{8}{c}{CP(\%)}\\
      \cmidrule(lr){3-4}\cmidrule(lr){6-7}\cmidrule(lr){9-16}\cmidrule(lr){18-25}
&& LP & IS && LP & IS && \multicolumn{2}{c}{MB} & \multicolumn{2}{c}{IS} & \multicolumn{2}{c}{SH} & \multicolumn{2}{c}{ZL} &&  \multicolumn{2}{c}{MB} & \multicolumn{2}{c}{IS} & \multicolumn{2}{c}{SH} & \multicolumn{2}{c}{ZL}\\
     \cmidrule(lr){8-10}\cmidrule(lr){11-12}\cmidrule(lr){13-14}\cmidrule(lr){15-17}\cmidrule(lr){18-19}\cmidrule(lr){20-21}\cmidrule(lr){22-23}\cmidrule(lr){24-25}
     &&& && &&& LP & IS& CF& MB & CF & MB & CF & MB && LP & IS& CF & MB & CF & MB & CF & MB\\
      \midrule
      \multicolumn{25}{l}{case cohort size = 150:}\\
      N & $\beta_1$ & 1.009 & 1.021 &  & 0.315 & 0.317 &  & 0.281 & 0.281 & 0.280 & 0.304 & 0.268 & 0.293 & 0.280 & 0.304 &  & 91.5 & 91.6 & 90.5 & 93.6 & 89.3 & 92.4 & 90.6 & 92.8 \\
      & $\beta_2$ & 1.014 & 1.029 &  & 0.172 & 0.172 &  & 0.151 & 0.150 & 0.149 & 0.160 & 0.145 & 0.157 & 0.149 & 0.161 &  & 91.9 & 92.1 & 90.8 & 93.5 & 89.3 & 91.0 & 90.8 & 93.1 \\
      & $\beta_3$ & 1.015 & 1.030 &  & 0.180 & 0.180 &  & 0.151 & 0.149 & 0.148 & 0.160 & 0.178 & 0.189 & 0.148 & 0.160 &  & 88.4 & 88.5 & 88.8 & 91.1 & 92.6 & 94.8 & 88.9 & 91.1 \\
      L & $\beta_1$ & 1.031 & 1.036 &  & 0.466 & 0.467 &  & 0.472 & 0.473 & 0.463 & 0.479 & 0.452 & 0.469 & 0.463 & 0.479 &  & 94.7 & 95.0 & 94.1 & 95.0 & 93.9 & 93.9 & 94.1 & 94.9 \\
      & $\beta_2$ & 1.032 & 1.039 &  & 0.254 & 0.254 &  & 0.256 & 0.256 & 0.250 & 0.257 & 0.252 & 0.261 & 0.250 & 0.258 &  & 95.0 & 95.0 & 94.5 & 95.1 & 95.1 & 95.5 & 94.4 & 95.3 \\
      & $\beta_3$ & 1.032 & 1.039 &  & 0.271 & 0.271 &  & 0.255 & 0.255 & 0.250 & 0.258 & 0.276 & 0.284 & 0.250 & 0.258 &  & 94.0 & 94.2 & 93.0 & 94.1 & 95.6 & 95.9 & 92.7 & 94.3 \\
      G & $\beta_1$ & 0.995 & 1.007 &  & 0.291 & 0.293 &  & 0.234 & 0.235 & 0.238 & 0.262 & 0.229 & 0.253 & 0.238 & 0.262 &  & 86.6 & 87.0 & 85.6 & 88.4 & 83.8 & 87.9 & 85.4 & 88.6 \\
      & $\beta_2$ & 1.015 & 1.030 &  & 0.152 & 0.153 &  & 0.129 & 0.128 & 0.128 & 0.139 & 0.123 & 0.134 & 0.128 & 0.139 &  & 90.0 & 90.5 & 88.4 & 91.4 & 85.9 & 89.0 & 88.5 & 91.8 \\
      & $\beta_3$ & 1.013 & 1.028 &  & 0.149 & 0.149 &  & 0.129 & 0.128 & 0.128 & 0.139 & 0.152 & 0.162 & 0.128 & 0.139 &  & 92.2 & 91.5 & 89.7 & 92.8 & 93.7 & 96.1 & 89.7 & 92.9 \\
      [1ex]
      \multicolumn{25}{l}{case cohort size = 300:}\\
      N & $\beta_1$ & 1.002 & 1.008 &  & 0.228 & 0.229 &  & 0.214 & 0.215 & 0.215 & 0.221 & 0.205 & 0.210 & 0.215 & 0.221 &  & 92.9 & 93.8 & 92.8 & 93.8 & 91.3 & 92.1 & 92.8 & 93.9 \\
      & $\beta_2$ & 1.005 & 1.012 &  & 0.130 & 0.130 &  & 0.118 & 0.118 & 0.119 & 0.122 & 0.111 & 0.114 & 0.119 & 0.122 &  & 92.8 & 93.6 & 93.3 & 94.6 & 91.2 & 92.7 & 93.4 & 94.0 \\
      & $\beta_3$ & 0.998 & 1.005 &  & 0.129 & 0.129 &  & 0.118 & 0.118 & 0.118 & 0.122 & 0.144 & 0.148 & 0.118 & 0.122 &  & 92.1 & 92.6 & 92.6 & 93.3 & 97.0 & 97.0 & 93.0 & 92.9 \\
      L & $\beta_1$ & 1.048 & 1.050 &  & 0.394 & 0.395 &  & 0.373 & 0.373 & 0.368 & 0.373 & 0.362 & 0.366 & 0.369 & 0.373 &  & 93.7 & 93.8 & 93.2 & 93.5 & 92.7 & 92.9 & 93.2 & 93.9 \\
      & $\beta_2$ & 1.031 & 1.035 &  & 0.221 & 0.222 &  & 0.205 & 0.205 & 0.204 & 0.207 & 0.204 & 0.207 & 0.204 & 0.207 &  & 92.1 & 92.9 & 92.6 & 93.0 & 92.9 & 92.9 & 92.6 & 92.8 \\
      & $\beta_3$ & 1.026 & 1.029 &  & 0.210 & 0.210 &  & 0.202 & 0.202 & 0.202 & 0.204 & 0.219 & 0.221 & 0.202 & 0.204 &  & 93.8 & 93.6 & 93.2 & 93.6 & 94.5 & 95.1 & 93.1 & 93.3 \\
      G & $\beta_1$ & 1.010 & 1.016 &  & 0.188 & 0.189 &  & 0.179 & 0.179 & 0.181 & 0.186 & 0.171 & 0.177 & 0.181 & 0.187 &  & 93.2 & 92.9 & 93.4 & 94.4 & 91.2 & 92.7 & 93.2 & 94.3 \\
      & $\beta_2$ & 1.007 & 1.015 &  & 0.107 & 0.108 &  & 0.099 & 0.098 & 0.099 & 0.102 & 0.092 & 0.095 & 0.099 & 0.102 &  & 91.9 & 92.6 & 92.4 & 93.3 & 90.1 & 91.4 & 92.3 & 93.0 \\
      & $\beta_3$ & 1.009 & 1.017 &  & 0.109 & 0.109 &  & 0.100 & 0.099 & 0.100 & 0.103 & 0.124 & 0.127 & 0.100 & 0.103 &  & 92.4 & 92.0 & 92.4 & 93.4 & 96.6 & 97.1 & 92.2 & 93.4 \\
      \bottomrule
    \end{tabular}
  \end{center}
\end{sidewaystable}

\begin{table}[tbp]
  \caption{Summary of timing results in seconds with both point estimation and
    variance estimation from the simulation study.}
  \label{tab:time}
  \begin{center}
    \begin{tabular}{rrr rr rrrrrrrrr}
      \midrule
      $C_p$ & $\bar m$ & Error &\multicolumn{2}{c}{PE}&\multicolumn{8}{c}{Variance}\\
      \cmidrule(lr){4-5}\cmidrule(lr){7-14}
      &&& LP & IS && \multicolumn{2}{c}{MB} & \multicolumn{2}{c}{IS} & \multicolumn{2}{c}{SH} & \multicolumn{2}{c}{ZL}\\
      \cmidrule(lr){7-8}\cmidrule(lr){9-10}\cmidrule(lr){11-12}\cmidrule(lr){13-14}
     &&&& &&  LP & IS& CF& MB & CF & MB & CF & MB\\
      \cmidrule(lr){1-14}
      \multicolumn{4}{l}{Full cohort size = 1500}\\
      90\% & 300 & N & 7.2 & 1.6 && 2007.5 & 561.3 & 2.9 & 10.9 & 2.9 & 11.4 & 2.1 & 11.6 \\
      && L & 6.5 & 1.5 && 1708.0 & 499.9 & 2.8 & 10.2 & 2.9 & 10.8 & 2.2 & 10.9 \\
      && G & 6.9 & 1.6 && 1899.7 & 544.6 & 2.8 & 10.4 & 2.8 & 10.9 & 2.0 & 11.1 \\
      [1ex]
      \multicolumn{4}{l}{Full cohort size = 3000}\\
      97\% & 150 & N & 0.8 & 0.6 && 183.4 & 150.2 & 0.4 & 2.8 & 0.5 & 3.0 & 0.6 & 3.0 \\
      &&  L & 0.7 & 0.4 && 143.7 & 118.2 & 0.3 & 2.5 & 0.5 & 2.6 & 0.6 & 2.7 \\
      &&  G & 1.1 & 0.7 && 262.3 & 191.6 & 0.5 & 3.5 & 0.7 & 3.7 & 0.7 & 3.7 \\
      [1ex]
      & 300 &  N & 3.6 & 0.9 && 629.6 & 301.7 & 2.1 & 5.5 & 2.2 & 5.8 & 1.3 & 5.8 \\
      &&  L & 3.3 & 0.7 && 544.9 & 237.1 & 2.0 & 4.8 & 2.1 & 5.1 & 1.3 & 5.2 \\
      &&  G & 4.1 & 1.2 && 816.8 & 367.8 & 2.2 & 6.5 & 2.3 & 6.8 & 1.4 & 6.9 \\
      \bottomrule
    \end{tabular}
  \end{center}
\end{table}

Results for the mildly rare disease case with censoring percentage
$C_p = 90\%$, full cohort size 1500, and average case-cohort size
$\bar m = 300$ are summarized in Table~\ref{tab:1500:90}.
Both the LP and the IS estimators appear to be virtually unbiased.
In fact, they agreed with each other closely on a 45 degree line (not shown).
Consequently, their empirical standard errors agreed with each other,
and their bootstrap based standard errors agreed with each other.
The bootstrap standard errors and the empirical standard errors
match closely, suggesting that the bootstrap variance estimators
provide good estimation of the empirical variantion.
The other six standard errors based on sandwich variance estimators
agreed quite well with the empirical standard errors too.
The associated 95\% confidence intervals based on all eight standard errors
had empirical coverage percentages reasonably close to the nominal level.
These observations were invariant to the error distributions.

Table~\ref{tab:3000:97} summarizes the results for the very rare
disease case with censoring rate 97\% and full cohort size 3000.
The results for full cohort size 1500 were similar and not reported.
The two point estimates, their empirical standard errors, and
their average bootstrap standard errors still agree with each other.
The bootstrap standard errors for case-cohort size 150, however, are
underestimating the true variation, and as a result, the 95\% confidence
intervals had coverage percentage smaller than the nominal level.
Not surprisingly, the six sandwich variance estimators performed
no better than the two multiplier bootstrap variance estimators.
When the case-cohort size was increased to 300, all variance estimators
performed reasonably well in estimating the true variation and
the coverage percentage was reasonably close to the nominal level.
Among all the sandwich variance estimators, the IS-MB and ZL-MB approaches
seem to provide confidence intervals with the best coverage percentage.
The SH-MB approach is slight inferior, which might be explained by
the fact that this approach has two layers of approximation ---
one from asymptotic linear approximation and the other from
induced smoothing.

Of more interest is Table~\ref{tab:time}, which summarizes the
timing results in seconds averaged from 1000 replicates for both
point estimation and variance estimation on a 2GHz linux machine.
For point estimation with full cohort size 1500 and censoring
percentage $C_p = 0.90\%$, the IS approach was up to 4.5 times
as fast as the LP approach (with normal error distribution).
The multipler bootstrap variance estimation with the IS approach
was up to 3.6 times as fast as the LP approach (again with normal error).
Nevertheless, the multiplier bootstrap IS approach still needed about
9 minutes on average to obtain a variance estimator.
All sandwich variance estimators are strikingly much faster,
especially with the closed-form approach:
ZL-CF approach took about 2 seconds on average;
the IS-CF and SH-CF approaches took about 3 seconds on average.
For each sandwich variance estimator, the version with CF estimation
of $V$ is over 5 times faster than the version with MB estimation of $V$.
Using the LP approach as benchmark, the IS-CF and SH-CF estimators
is 695 times faster and the ZL-CF estimators 1003 times faster.
Since the performance of all variance estimators are similar for
this setting, the IS-CF, SH-CF and ZL-CF approaches are obviously
preferred for this setup with a mildly rare event.

The timing results for full cohort size 3000 and censoring
percentage $C_p = 0.97\%$ follow a similar pattern.
Compare to case $C_p = 0.90\%$, time for point estimation is shorter
because the number of cases decreases in the case $C_p = 0.97\%$
even when the average case-cohort size were both at 300.
Sandwich variance estimators with CF estimation of $V$ is up to
8 times faster than with those with MB estimation of $V$, at the
expense of slightly worse performance in coverage percentage.
The IS-MB and ZL-MB approaches yield the most reliable variance
estimates but IS-MB is slightly faster than faster than ZL-MB.
As the average case-cohort size doubles, the computing time
of the sandwich variance estimates with MB estimation for $V$
appear to double accordingly, in contrast to those with CF
estimation for $V$, which do not necessarily double linearly.
In summary, based on the performance and speed, our recommended
variance estimator is the IS-MB estimator.

\section{National Wilm's Tumor Study}
\label{sect:appl}

We demonstrate the performance of our proposed methods with an
application to the cohort study conducted by the National Wilm's
Tumor Study Group (NWTSG) \citep{DAng:trea:1989, Gree:comp:1998}.
Wilm's tumor is a rare kidney cancer in young children.
The interest of the study was to assess the relationship between
the tumor histology and the outcome, time to tumor relapse.
Tumor histology can be classified into two categories,
favorable or unfavorable, depending on the cell type.
The central histological diagnosis was made by an individual
pathologist at the central pathology center, which was
believed to be more accurate than a local diagnosis yet
more expensive to measure and required more efforts to obtain.
Although in the full version of the data, the central histology 
measurement was available for all the cohort members, it was only
available for a case-cohort sample in the case-cohort version.
We take advantage of the full version in this example.
Other covariates that were available for all cohort
members were patient age, disease stage and study group.
According to the staging system employed by NWTSG,
four stages (I -- IV) of Wilms' tumors, with Stage IV as the
latest stage, indicated the spread of the tumor.
Each subject came from one of the two study groups, NWTSG-3 and NWTSG-4.
The case-cohort version of the data was analyzed with 
Cox models \citep{Kuli:Lin:impr:2004, Bres:Luml:Ball:impr:2009} 
and additive hazards models \citep{Kuli:Lin:addi:2000}, respectively.

There were a total of 4028 subjects in the full cohort.
Among them, 571 were cases who experienced the relapse
of tumor --- a censoring rate of about 86\%.
We considered an AFT model for the time to relapse with
the following covariates:
central histology measurement (1 = favorable, 0 = unfavorable),
age (measure in year) at diagnosis,
three tumor stages indicators (Stage I as reference)
and a study group indicator (NWTSG-3 as reference).
The case-cohort version of the data had 668 patients selected as
sub-cohort sample and the total case-cohort sample size was 1154.
To take advantage of availability of full cohort data,
we drew 1000 new sub-cohort samples with size 668 and formed
a case-cohort by including the remaining cases for each replicate.
We then averaged these estimates and estimated standard
errors from the 1000 replicates of the case-cohort analysis.

\begin{table}[tbp]
  \caption{National Wilm's tumor study and timing results in seconds.}
  \label{tab:tumo}
  \begin{center}
    \begin{tabular}{c rc crrrrrrr}
      \toprule
      &\multicolumn{1}{r}{PE}& &\multicolumn{7}{c}{SE}\\
      \cmidrule(lr){2-3} \cmidrule(lr){4-10}
      Effects& IS & & MB & \multicolumn{2}{c}{IS} & \multicolumn{2}{c}{SH} & \multicolumn{2}{c}{ZL}\\
	\cmidrule(lr){4-4}\cmidrule(lr){5-6}\cmidrule(lr){7-8}\cmidrule(lr){9-10}
      &&&IS&CF&MB&CF&MB&CF&MB\\
      \midrule
      \multicolumn{10}{l}{Case-Cohort Analysis:}\\
      (time) & (8.5) && (3682.2) & (7.7) & (13.9) & (9.5) & (15.7) & (9.4) & (15.0) \\
      histol &$-$3.428 && 0.465 & 0.409 & 0.458 & 0.372 & 0.423 & 0.410 & 0.458 \\
      age & $-$0.190 && 0.079 & 0.074 & 0.080 & 0.221 & 0.243 & 0.074 & 0.080 \\
      stage2 & $-$1.283 && 0.613 & 0.590 & 0.621 & 0.516 & 0.544 & 0.590 & 0.622 \\
      stage3 & $-$1.401 && 0.612 & 0.579 & 0.616 & 0.572 & 0.602 & 0.580 & 0.616 \\
      stage4 & $-$2.092 && 0.717 & 0.665 & 0.712 & 0.717 & 0.763 & 0.666 & 0.712 \\
      study & $-$0.128 && 0.475 & 0.455 & 0.484 & 0.451 & 0.482 & 0.455 & 0.484 \\
      [1ex]
      \multicolumn{10}{l}{Full-Cohort Analysis:}\\
      (time) &(266.0)&& (126927.7) & (309.9) & (453.0) & (341.3) & (486.1) & (321.1) & (494.0) \\
      histol & $-$2.749 && 0.202 & 0.148 & 0.213 & 0.138 & 0.214 & 0.148 & 0.196 \\
      age & $-$0.127 && 0.037 & 0.029 & 0.039 & 0.081 & 0.092 & 0.029 & 0.039 \\
      stage2 & $-$1.335 && 0.280 & 0.233 & 0.285 & 0.200 & 0.280 & 0.234 & 0.271 \\
      stage3 & $-$1.341 && 0.286 & 0.239 & 0.297 & 0.211 & 0.288 & 0.240 & 0.299 \\
      stage4 & $-$2.203 && 0.319 & 0.245 & 0.321 & 0.219 & 0.300 & 0.247 & 0.334 \\
      study & $-$0.106 && 0.226 & 0.175 & 0.229 & 0.162 & 0.224 & 0.176 & 0.219 \\
      \bottomrule
    \end{tabular}
  \end{center}
\end{table}

The results of the average from 1000 replicates of
case-cohort analyses are summarized in Table~\ref{tab:tumo}.
Due to its poor timing performance, the LP approach was not considered.
Since the MB standard error is considered to reflect the true variation
quite well from the simulation study, we are interested in how close
the various sandwich standard errors to the MB standard error.
For all three sandwich estimators, the CF versions systematically
underestimate noticeably, although the underestimation is less
severe in the IS and ZL approach than in the SH approach.
The MB versions of the sandwich estimates appear to agree with
the MB standard error closely, and again the agreement appears
to be better for the IS and ZL approach than for the SH approach.
In particular, the SH standard error for the age effect
is about three times as much as that from other approaches.
The standard errors from IS-MB and ZL-MB are almost identical,
both very close to the time consuming MB standard error.
Based on the IS-MB standard errors, the coefficients of central
histological diagnosis, age, and all three stage indicators were
found to be significantly different from zero with p-values
0.000, 0.009, 0.019, 0.011 and 0.002, respectively.
No significant difference was found between the two study groups.
In terms of timing, the MB-IS standard error took over a hour
whereas the MB-based sandwich estimates only took 14--15 seconds on average.

For comparison purpose, we also analyzed the full cohort data with
the same approaches and reported the results in Table~\ref{tab:tumo}.
Point estimates are close to these in case-cohort analysis,
with their standard errors taken into consideration.
All the standard errors decrease compared to the case-cohort analyses,
which is expected as full information became available for all covariates.
The best sandwich variance estimators are still IS-MB and ZL-MB,
both closely approximates the full blown MB standard error.
With full cohort size 4028 and censoring rate 86\%, the IS point
estimates took 4 and a half minutes, the MB variance estimation
took 35.36 hours, while IS-MB only took 7 and a half minutes.

\section{Discussion}
\label{sect:disc}

In AFT modeling of case-cohort data, both point estimation and variance
estimation are challenging with the nonsmooth estimating equations.
Resampling methods are commonly used to estimate the variance, which
are time consuming even with a computationally efficient point estimator
such as our induced smoothing approach with rank-based estimating equations.
We have proposed six sandwich variance estimators and compared their
performances with the bootstrap variance estimator in numerical studies.
The IS-MB and ZL-MB approaches were found to provide good approximation
to the true variation and are computationally very efficient.
All the methods are implemented in an R package aftgee \citep{Rpkg:aftgee}.
The package had the potential to bring AFT modeling of case-cohort
data into routine analysis.

The IS approach was built on Gehan's weight for rank-based
estimating equations, in which case closed-form expectations
of the perturbed estimating equations are available.
Alternative weights such as the logrank weight are possible, though the
computation is less straightforward than that for Gehan's weight.
Incorporating a general, possibly optimized weight in the IS approach
merits further investigation for both full cohort and case-cohort data.
The estimates from Gehan's weight always serve as a good initial
value in numerical equations solving.

Some extensions of the proposed methods are worth considering.
Auxiliary covariates are often available for the entire cohort,
which can be used to construct strata for subcohort members selection.
The resulting estimators from the stratified case-cohort design 
has been shown to be more efficient than their traditional case-cohort 
counterpart for the Cox model \citep{Kuli:Lin:impr:2004} 
and the additive hazards model \citep{Kuli:Lin:addi:2000}. 
An extension of the proposed methods with the AFT model to a 
stratified case-cohort design may lead to efficiency improvement too.
When more than one diseases are considered in a case-cohort design,
a multivariate extension will be needed, and a possible dependence 
among the multivariate failure times needs to be taken into account. 
For the Cox model, \citet{Kang:Cai:marg:2009} used a marginal approach.
A similar approach can be considered for the AFT model.

\appendix

\section{Analytical Details}
We give the analytical form of $S_i(\beta)$'s here.
Define the general rank based weighted estimating function
\citep{Jin:Lin:Wei:Ying:rank:2003}
\begin{equation*}
  U_n(\beta)=\sum_{i=1}^n \Delta_i \varphi_{n,i}(\beta)\left[ X_i-\frac{W^{(1)}_{n,i}(\beta)}{W^{(0)}_{n,i}(\beta)}\right],
\end{equation*}
where $\varphi_{n,i}(\beta)$ is an nonnegative weight function and
\begin{equation*}
  W^{(k)}_{n,i}(\beta)=\frac{1}{n}\sum_{j=1}^n X_j^k I[e_j(\beta) \geq e_i(\beta)],
  \qquad k = 0,1.
\end{equation*}
Equation~\eqref{eq:Un} can be obtained by setting
$\varphi_{n,i}(\beta) = W^{(0)}_{n,i}(\beta)$.
On the other hand,
the general rank based weighted estimating function for case-cohort
samples has the following form:
\begin{equation*}
  U_n^c(\beta) = \sum_{i=1}^n\Delta_i\varphi_{n,i}(\beta)\left[X_i-\frac{\hat{W}^{(1)}_{n, i}(\beta)}{\hat{W}^{(0)}_{n, i}(\beta)}\right],
\end{equation*}
where
\begin{equation*}
  \hat{W}^{(k)}_{n, i}(\beta)=\frac{1}{n} \sum_{j=1}^n h_j X_j^k I[e_j(\beta)\geq e_i(\beta)], \qquad k = 0,1.
\end{equation*}
Similarly, equation~\eqref{eq:UnC} can be obtained by setting $\varphi_{n,i}(\beta) =
\hat{W}^{(0)}_{n,i}(\beta)$.

With these settings, an explicit form of $S_i(\beta_0)$ is
\begin{align*}
  S_i(\beta_0) =& \int_{-\infty}^\infty w^{(0)}(\beta_0)\left[ X_i-\frac{w^{(1)}(\beta_0)}{w^{(0)}(\beta_0)}\right] \, \dif M_i(t) \\
  =& \Delta_i w^{(0)}(\beta_0)\left[ X_i-\frac{w^{(1)}(\beta_0)}{w^{(0)}(\beta_0)}\right]-\int_{-\infty}^{e_i(\beta)} w^{(0)}(\beta_0)\left[ X_i-\frac{w^{(1)}(\beta_0)}{w^{(0)}(\beta_0)}\right] \lambda(t)\,\dif t,
\end{align*}
where
\begin{equation*}
  w^{(k)}(\beta)=\lim_{n\to \infty} \hat{W}^{(k)}_{n, i}(\beta), \mbox{ for }k = 0,1,
\end{equation*}
\begin{equation*}
  M_i(t)=N_i(\beta;t)-\int_0^t I(e_i(\beta)\geq u) \lambda(u)\, \dif u,
\end{equation*}
$N_i(\beta;t) = \Delta_iI(e_i(\beta)\le t)$ and
$\lambda(u)$ is the common hazard function of $\epsilon_i$.

The unknown quantities in $S_i(\beta_0)$ include
$\beta_0$, $w^{(0)}$, $w^{(1)}$ and $\lambda(t)$.
With the explicit form of $S_i(\beta_0)$,
$\hat{S}_i(\hat \beta)$ is obtained by replacing these
unknown quantities by their sample estimators.

% \bibliographystyle{stco}
% \bibliography{cohort}

\begin{thebibliography}{34}
\expandafter\ifx\csname natexlab\endcsname\relax\def\natexlab#1{#1}\fi
\expandafter\ifx\csname url\endcsname\relax
  \def\url#1{{\tt #1}}\fi
\expandafter\ifx\csname urlprefix\endcsname\relax\def\urlprefix{URL }\fi

\bibitem[{Barlow(1994)}]{Barl:robu:1994}
Barlow, W.~E. 1994. Robust variance estimation for the case-cohort design.
  Biometrics 50: 1064--1072.

\bibitem[{Breslow et~al.(2009)Breslow, Lumley, Ballantyne, Chambless, and
  Kulich}]{Bres:Luml:Ball:impr:2009}
Breslow, N.~E., Lumley, T., Ballantyne, C.~M., Chambless, L.~E., and Kulich, M.
  2009. Improved horvitz--thompson estimation of model parameters from
  two-phase stratified samples: Applications in epidemiology. Statistics in
  Biosciences 1: 32--49.

\bibitem[{Brown and Wang(2005)}]{Brow:Wang:stan:2005}
Brown, B.~M. and Wang, Y.-G. 2005. Standard errors and covariance matrices for
  smoothed rank estimators. Biometrika 92: 149--158.

\bibitem[{Brown and Wang(2007)}]{Brow:Wang:indu:2007}
Brown, B.~M. and Wang, Y.-G. 2007. Induced smoothing for rank regression with
  censored survival times. Statistics in Medicine 26: 828--836.

\bibitem[{Chen(2001{\natexlab{a}})}]{Chen:fitt:2001}
Chen, H.~Y. 2001{\natexlab{a}}. Fitting semiparametric transformation
  regression models to data from a modified case-cohort design. Biometrika 88:
  255--268.

\bibitem[{Chen(2001{\natexlab{b}})}]{Chen:weig:2001}
Chen, H.~Y. 2001{\natexlab{b}}. Weighted semiparametric likelihood method for
  fitting a proportional odds regression model to data from the case cohort
  design. Journal of the American Statistical Association 96: 1446--1458.

\bibitem[{Chiou et~al.(2012)Chiou, Kang, and Yan}]{Rpkg:aftgee}
Chiou, S., Kang, S., and Yan, J. 2012. {aftgee}: {A}ccelerated Failure Time
  Model with Generalized Estimating Equations. R package version 0.2-27.

\bibitem[{D{'A}ngio et~al.(1989)D{'A}ngio, Breslow, Beckwith, Evans, Baum,
  Delorimier, Fernbach, Hrabovsky, Jones, Kelalis, Othersen, Tefft, and
  Thomas}]{DAng:trea:1989}
D{'A}ngio, G.~J., Breslow, N., Beckwith, J.~B., Evans, A., Baum, E.,
  Delorimier, A., Fernbach, D., Hrabovsky, E., Jones, B., Kelalis, P.,
  Othersen, H.~B., Tefft, M., and Thomas, P. R.~M. 1989. Treatment of {W}ilms'
  tumor. results of the third national {W}ilms' tumor study. Cancer 64:
  349--360.

\bibitem[{Green et~al.(1998)Green, Breslow, Beckwith, Finklestein, Grundy,
  Thomas, Kim, Shochat, Haase, Ritchey, Kelalis, and
  D{'A}ngio}]{Gree:comp:1998}
Green, D., Breslow, N., Beckwith, J., Finklestein, J., Grundy, P., Thomas, P.,
  Kim, T., Shochat, S., Haase, G., Ritchey, M., Kelalis, P., and D{'A}ngio, G.
  1998. Comparison between single-dose and dvided-dose administration of
  dactinomycin and doxorubicin for patients with {W}ilms' tumor: {A} report
  from the {N}ational {W}ilms' {T}umor {S}tudy {G}roup. Journal of Clinical
  Oncology 16: 237--245.

\bibitem[{Hasselman(2012)}]{Rpkg:nleqslv}
Hasselman, B. 2012. {nleqslv}: {S}olve systems of non linear equations. R
  package version 1.9.3.
\newline\urlprefix\url{http://CRAN.R-project.org/package=nleqslv}

\bibitem[{Huang(2002)}]{Huan:cali:2002}
Huang, Y. 2002. Calibration regression of censored lifetime medical cost.
  Journal of the American Statistical Association 97: 318--327.

\bibitem[{Jin et~al.(2003)Jin, Lin, Wei, and Ying}]{Jin:Lin:Wei:Ying:rank:2003}
Jin, Z., Lin, D.~Y., Wei, L.~J., and Ying, Z. 2003. Rank-based inference for
  the accelerated failure time model. Biometrika 90: 341--353.

\bibitem[{Johnson and Strawderman(2009)}]{John:Stra:indu:2009}
Johnson, L.~M. and Strawderman, R.~L. 2009. Induced smoothing for the
  semiparametric accelerated failure time model: {A}symptotics and extensions
  to clustered data. Biometrika 96: 577--590.

\bibitem[{Kalbfleisch and Lawless(1988)}]{Kalb:Lawl:like:1988}
Kalbfleisch, J.~D. and Lawless, J.~F. 1988. Likelihood analysis of multistate
  models for disease incidence and mortality. Statistics in Medicine 7:
  149--160.

\bibitem[{Kang and Cai(2009)}]{Kang:Cai:marg:2009}
Kang, S. and Cai, J. 2009. Marginal hazards model for case-cohort studies with
  multiple disease outcomes. Biometrika 96: 887--901.

\bibitem[{Kong and Cai(2009)}]{Kong:Cai:case:2009}
Kong, L. and Cai, J. 2009. Case-cohort analysis with accelerated failure time
  model. Biometrics 65: 135--142.

\bibitem[{Kong et~al.(2004)Kong, Cai, and Sen}]{Kong:Cai:Sen:weig:2004}
Kong, L., Cai, J., and Sen, P.~K. 2004. Weighted estimating equations for
  semiparametric transformation models with censroed data from a case-cohort
  design. Biometrika 91: 305--319.

\bibitem[{Kulich and Lin(2000)}]{Kuli:Lin:addi:2000}
Kulich, M. and Lin, D. 2000. Additive hazards regression for case-cohort
  studies. Biometrika 87: 73--87.

\bibitem[{Kulich and Lin(2004)}]{Kuli:Lin:impr:2004}
Kulich, M. and Lin, D. 2004. Improving the efficiency of relative-risk
  estimation in case-cohort studies. Journal of the American Statistical
  Association 99: 832--844.

\bibitem[{Lin and Ying(1993)}]{Lin:Ying:cox:1993}
Lin, D.~Y. and Ying, Z. 1993. Cox regression with incomplete covariate
  measurements. Journal of the American Statistical Association 88: 1341--1349.

\bibitem[{Lu and Tsiatis(2006)}]{Lu:Tsia:semi:2006}
Lu, W. and Tsiatis, A.~A. 2006. Semiparametric transformation models for the
  case-cohort study. Biometrika 93: 207--214.

\bibitem[{Nan et~al.(2006)Nan, Yu, and Kalbfleisch}]{Nan:Yu:Kalb:cens:2006}
Nan, B., Yu, M., and Kalbfleisch, J.~D. 2006. Censored linear regression for
  case-cohort studies. Biometrika 93: 747--762.

\bibitem[{Prentice(1978)}]{Pren:line:1978}
Prentice, R.~L. 1978. Linear rank tests with right censored data ({C}orr: {V}70
  p304). Biometrika 65: 167--180.

\bibitem[{Prentice(1986)}]{Pren:case:1986}
Prentice, R.~L. 1986. A case-cohort design for epidemiologic cohort studies and
  disease prevention trials. Biometrika 73: 1--11.

\bibitem[{Self and Prentice(1988)}]{Self:Pren:asym:1988}
Self, S.~G. and Prentice, R.~L. 1988. Asymptotic distribution theory and
  efficiency results for case-cohort studies. The Annals of Statistics 16:
  64--81.

\bibitem[{Sun et~al.(2004)Sun, Sun, and Flournoy}]{Sun:Sun:Flou:addi:2004}
Sun, J., Sun, L., and Flournoy, N. 2004. Additive hazards model for competing
  risks analysis of the case-cohort design. Communications in Statistics:
  Theory and Methods 33: 351--366.

\bibitem[{Therneau and Li(1999)}]{Ther:Li:comp:1999}
Therneau, T.~M. and Li, H. 1999. Computing the cox model for case cohort
  designs. Lifetime Data Analysis 5: 99--112.

\bibitem[{Tsiatis(1990)}]{Tsia:esti:1990}
Tsiatis, A.~A. 1990. Estimating regression parameters using linear rank tests
  for censored data. The Annals of Statistics 18: 354--372.

\bibitem[{Varadhan and Gilbert(2009)}]{Vara:Gilb:BB:2009}
Varadhan, R. and Gilbert, P. 2009. {BB}: An {R} package for solving a large
  system of nonlinear equations and for optimizing a high-dimensional nonlinear
  objective function. Journal of Statistical Software 32: 1--26.
\newline\urlprefix\url{http://www.jstatsoft.org/v32/i04/}

\bibitem[{Wacholder et~al.(1989)Wacholder, Gail, Pee, and
  Brookmeyer}]{Wach:Gail:Pee:Broo:altn:1989}
Wacholder, S., Gail, M.~H., Pee, D., and Brookmeyer, R. 1989. Alternative
  variance and efficiency calculations for the case-cohort design. Biometrika
  76: 117--123.

\bibitem[{Ying(1993)}]{Ying:larg:1993}
Ying, Z. 1993. A large sample study of rank estimation for censored regression
  data. The Annals of Statistics 21: 76--99.

\bibitem[{Yu(2011)}]{Yu:buck:2011}
Yu, M. 2011. Buckley-james type estimator in censored data with covariates
  missing by design. Scandinavian Journal of Statistics 38: 252--267.

\bibitem[{Yu et~al.(2007)Yu, Wong, and Yu}]{Yu:Wong:Yu:buck:2007}
Yu, Q., Wong, G. Y.~C., and Yu, M. 2007. Buckley-{J}ames-type of estimators
  under the classical case cohort design. Annals of the Institute of
  Statistical Mathematics 59: 675--695.

\bibitem[{Zeng and Lin(2008)}]{Zeng:Lin:effi:2008}
Zeng, D. and Lin, D.~Y. 2008. Efficient resampling methods for nonsmooth
  estimating functions. Biostatistics 9: 355--363.

\end{thebibliography}

\end{document}